\documentclass{elsart}
\usepackage[english]{babel}
\usepackage{graphicx,subfigure}
\usepackage{cite}
\usepackage{amsmath,amssymb}

\begin{document}
\begin{frontmatter}

\title{%
 Dalitz plot analysis of the
 $D^+ \to K^- \pi^+ \pi^+$ decay in the FOCUS experiment.}


\collab{The~FOCUS~Collaboration\thanksref{AUTHORS}}

\thanks[AUTHORS]{See \textrm{http://www-focus.fnal.gov/authors.html} for additional
author information.}

\author[ucd]{J.~M.~Link}
\author[ucd]{P.~M.~Yager}
\author[cbpf]{J.~C.~Anjos}
\author[cbpf]{I.~Bediaga}
\author[cbpf]{C.~Castromonte}
\author[cbpf]{A.~A.~Machado}
\author[cbpf]{J.~Magnin}
\author[cbpf]{A.~Massafferri}
\author[cbpf]{J.~M.~de~Miranda}
\author[cbpf]{I.~M.~Pepe}
% ...
\null

\address[ucd]{University of California, Davis, CA 95616}
\address[cbpf]{Centro Brasileiro de Pesquisas F\'\i sicas, Rio de Janeiro, RJ, Brazil}
% ...
\address[fnal]{Fermi National Accelerator Laboratory, Batavia, IL 60510}
% ...


\begin{abstract}

Using data collected by the high energy photoproduction experiment FOCUS at
Fermilab we performed a Dalitz plot analysis of the Cabibbo favored decay
\mbox{$D^+\to K^- \pi^+ \pi^+$}. This study uses 53653 Dalitz-plot events with
a signal fraction of $\sim$ 97\%, and represents the highest statistics, most
complete Dalitz plot analysis for this channel.
Results are presented and discussed using two different formalisms. The first
is a simple sum of Breit--Wigner functions with freely fitted masses and
widths. It is the model traditionally adopted and serves as comparison with
the already published analyses. The second uses a {\it K-matrix} approach for
the dominant $S$-wave, in which the parameters are fixed by first fitting
$K\pi$ scattering data and continued to threshold by Chiral Perturbation
Theory. We show that the Dalitz plot distribution for this decay is consistent
with the assumption of two body dominance of the final state interactions and
the description of these interactions is in agreement with other data on the
$K\pi$ final state.
\end{abstract}

\end{frontmatter}


\section{Introduction}

In recent years, the Dalitz plot technique has been widely applied to the
heavy-flavor sector. It is acknowledged as a powerful tool with which to study
charm and beauty decay dynamics, and so test the consistency of the Standard
Model, investigate CP violation effects and set limits on new physics. To
perform precision studies requires an accurate description of the hadron
dynamics that colors and shapes the final states. Whether in the three-pion
\cite{K_matrix}, or here in the $K \pi\pi$ final state, two body interactions
with scalar quantum numbers dominate the Dalitz plot distribution.
Consequently broad overlapping resonances control the dynamics. A coherent sum
of Breit--Wigners can, in general, provide an adequate description of data.
However, the effective Breit--Wigner mass and width need not accurately
reflect the true positions of any resonance poles, particularly for wide
states like the $\kappa$ and $\sigma$. If three-body interactions play a
limited role, then we can adopt a parametrization which enforces the two-body
unitarity constraint and is consistent with the two-body scattering data. This
is naturally embodied in the \emph{K-matrix} formalism. The FOCUS
collaboration has already performed a pioneering Dalitz plot analysis of the
$D^+$ and $D^+_s \to \pi^+\pi^-\pi^+$ decays \cite{K_matrix} implementing the
\emph{K-matrix} formalism for the description of the $\pi \pi$ $S$-wave
intermediate states. It led us to conclude that the three pion $D$-decay is
well-described using data from $\pi\pi$ scattering and that any $\sigma$-like
object in $D$-decay is consistent with the $\sigma$ extracted from $\pi\pi$
scattering \cite{CCL}. The same conclusion was reached by other authors (see
for instance \cite{Bugg} and \cite{Oller}). The $D^+ \to K^-\pi^+\pi^+$
analysis discussed in this letter represents the analogous study in the $K\pi$
system.

It is worth noting that the \emph{K-matrix} formalism \cite{wigner,chung},
originating in the context of two-body scattering, can be generalized to cover
the case of production of resonances in more complex reactions \cite{aitch},
with the assumption that the two-body system in the final state is an isolated
one and that the two particles do not simultaneously interact with the rest of
the final state in the production process \cite{chung}. The validity of the
assumed quasi two-body nature of the process of the \emph{K-matrix} approach
can only be verified by a direct comparison of the model predictions with
data. In particular, the failure to reproduce the Dalitz plot distribution
could be an indication of the presence of relevant, neglected three-body
effects.

Within the $K$-matrix approach, we present the first determination in
$D$-decays of the two separate isospin contributions, $I = 1/2$ and $I=3/2$,
for the $S$-wave $K\pi$ system. The $I=1/2$ component is the most important,
being dominated by broad resonances. In comparison, the $I=3/2$ contribution
is small at low $K\pi$ masses, but becomes larger at higher masses. This
results in a significant interference between these components to describe the
full Dalitz plot distribution. We will see that our results indicate close
consistency with $K\pi$ scattering data, and consequently with Watson's
theorem predictions for two-body $K\pi$ interactions in the low $K\pi$ mass
region, where elastic processes dominate.


\section{Signal selection}

The data for this analysis were collected during the 1996--1997 run of the
photoproduction experiment FOCUS at Fermilab. The detector, designed and used
to study the interaction of high-energy photons on a segmented BeO target, is
a large aperture, fixed-target magnetic spectrometer with excellent
\v{C}erenkov particle identification and vertexing capabilities. Most of the
FOCUS experiment and analysis techniques have been described previously
\cite{E687_spectr,Focus_cherenkov,life_diff}.

The FOCUS collaboration has already published the analysis of the $D^+$ meson
lifetime in the $K^-\pi^+ \pi^+$ channel \cite{dpiu_lifetime}. Candidates for
the present Dalitz plot analysis have been selected according to an analogous
set of cuts. A decay vertex is formed from three reconstructed charged tracks.
The momentum of the $D$ candidate is used to intersect other reconstructed
tracks to form a production vertex. The confidence level (C.L.) of each vertex
is required to exceed 1\,\%. After the vertex finder algorithm, the variable
$\ell$, which is the separation of the primary and secondary vertices, and its
associated error $\sigma_\ell$ are calculated. We reduce backgrounds by
requiring $\ell/\sigma_\ell>15$. The two vertices are also required to satisfy
isolation conditions. The primary vertex isolation cut requires that a track
assigned to the decay vertex has  a C.L. less than 0.1\,\% to be included in
the primary vertex. The secondary vertex isolation cut requires that all
remaining tracks not assigned to the primary and secondary vertex have a C.L.\
smaller than 0.001\,\% to form a vertex with the $D$ candidate daughters.

\begin{figure}[!ht]
\centering
  {
 \includegraphics[width=0.45\textwidth]{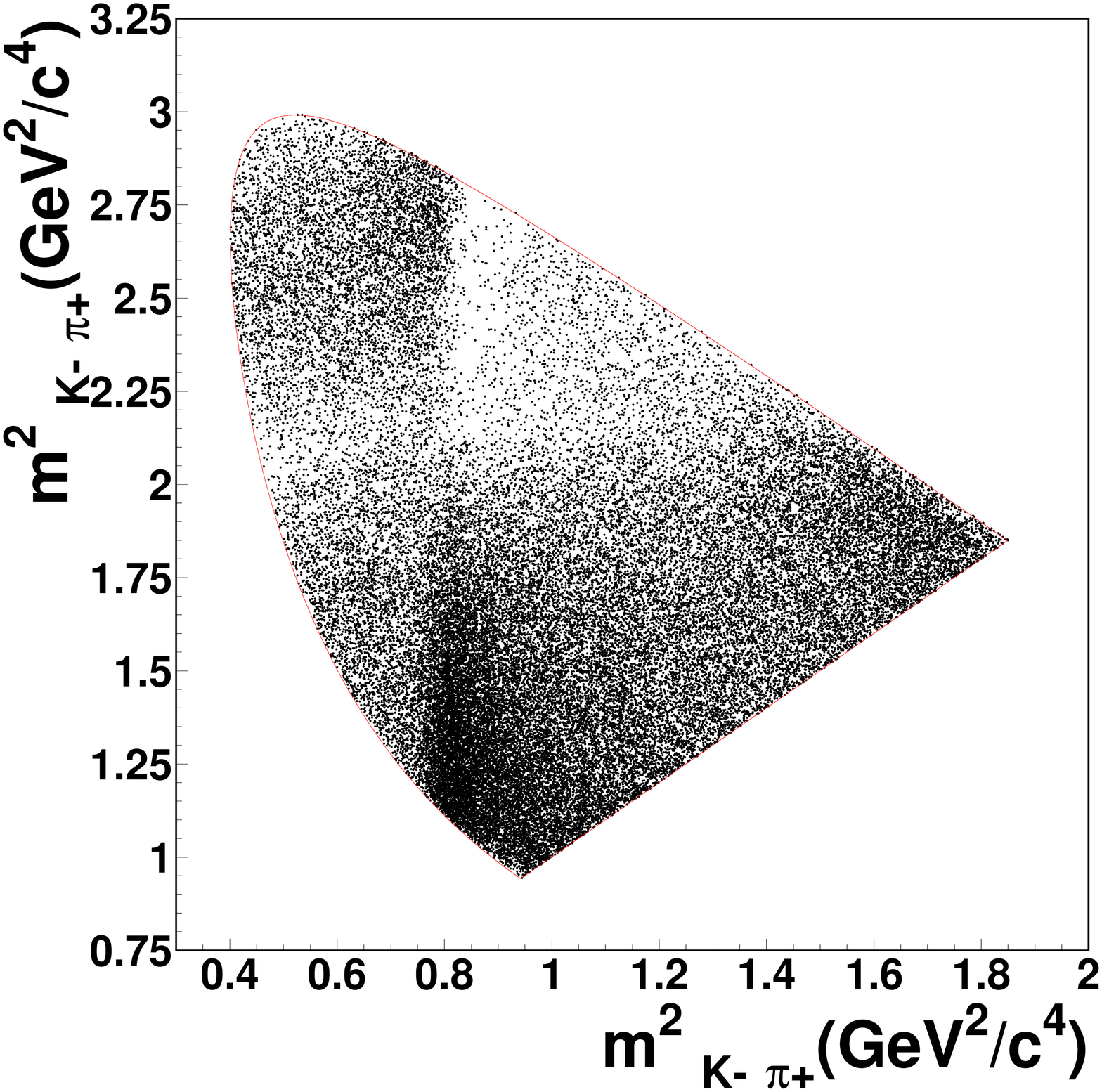}
 \includegraphics[width=0.45\textwidth]{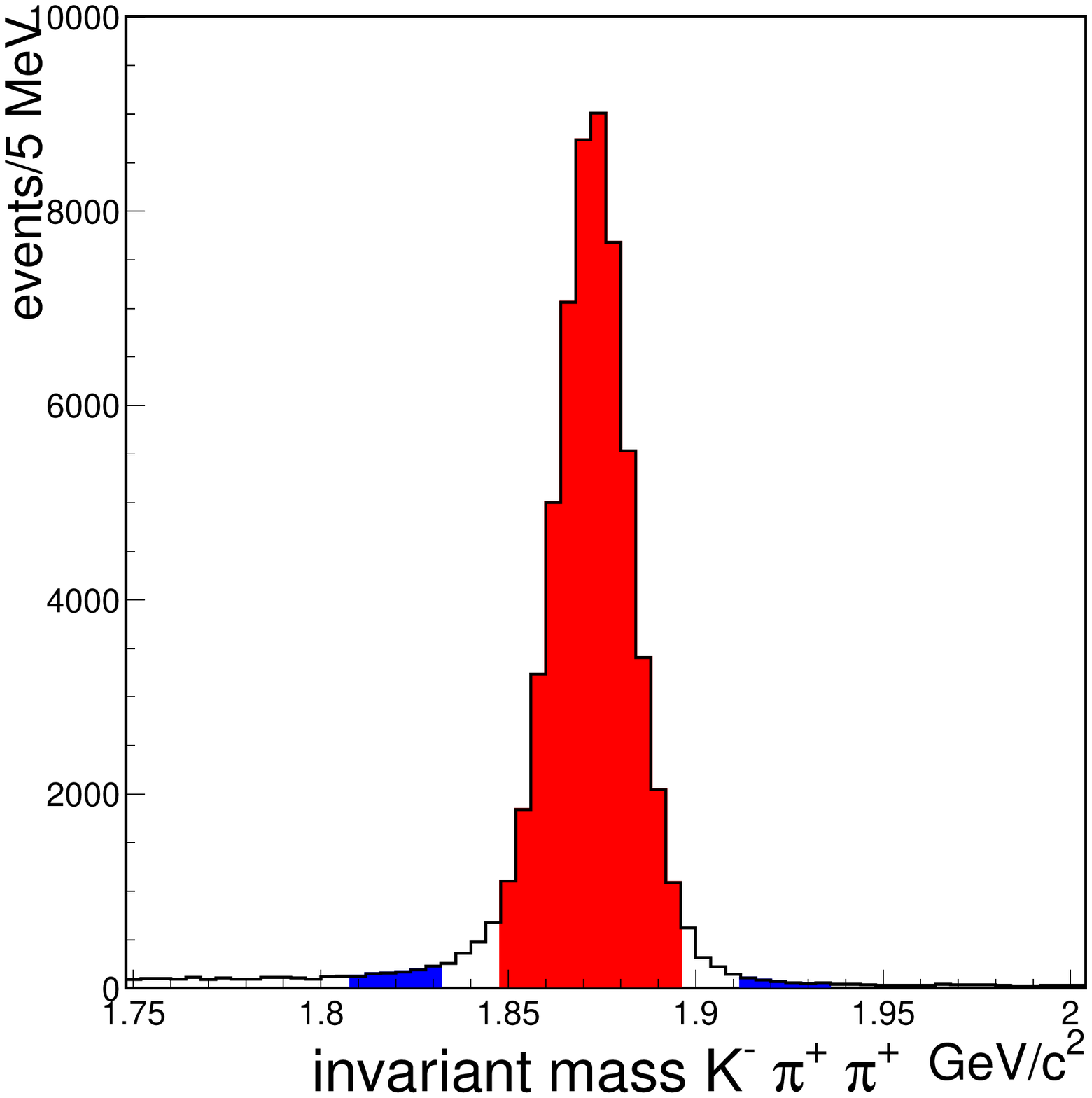}
}
 \caption{$D^+ \to K^-\pi^+\pi^+$ Dalitz plot and mass distribution: signal and sideband regions
 are indicated in red and blue respectively. Sideband are at $\pm$(6--8)
 $\sigma$ from the peak.}
\label{signal}
\end{figure}

The \v{C}erenkov  particle identification is based on likelihood ratios
between the various stable particle hypotheses \cite{Focus_cherenkov}. The
product of all firing probabilities for all cells within the three
\v{C}erenkov cones produces a $\chi^2$-like variable $W_i=-2\ln$(likelihood)
where $i$ ranges over the electron, pion, kaon and proton hypothesis. We
require $\Delta W_K= W_{\pi} -W_{K}$  $> 5$ for the kaon candidate, and
$\Delta W_{\pi}= W_{K} -W_{\pi}$  $> 5 $ for both pions in the final state.
Kaon and pion consistency, $\Delta W=W_{K}-W_{\min}<3$ and $\Delta
W=W_{\pi}-W_{\min}<3$ is also required where $W_{\min}$ is the minimum $W_i$.
These \v{C}erenkov cuts reduce the residual contamination of the $D_s^{\pm}
\to K^{\mp}K^{\pm}\pi^{\pm}$, when a kaon is mis-identified as a pion, to a
negligible level. The final sample invariant mass distribution is shown in
Fig.~\ref{signal}. Yields for signal and background, evaluated within $\pm 2
\sigma$ from the mass peak, i.e, from 1.85 to 1.89 GeV, consist of 52460 $\pm$
245 and 1897 $\pm$ 39 events, respectively. Events satisfying the kinematic
limit condition populate the final Dalitz plot, shown in the same
Fig.~\ref{signal}, on which the amplitude analysis is performed.


\section{The Dalitz plot analysis}

Previous analyses of the $D^+ \to K^-\pi^+\pi^+$ used an isobar model, where
the decay amplitude consists of a simple sum of Breit--Wigner functions.  In
this analysis, the masses and widths for scalar states will be determined by
fitting to this particular decay channel. This model usefully serves as the
standard for quality of fit, with no presumption as to the correctness of the
physics the model embodies. The limitation of this approach, as already
mentioned, is that the parameters of the scalar states are ``ad hoc'',
established with no reference to those found in other processes, in particular
scattering data. Moreover, the states involved are wide and overlapping and
the Breit--Wigner modelling is too simplistic. Assuming dominance of two-body
interactions the {\it K-matrix} formalism \cite{wigner,chung} provides the
correct treatment for the $S$-wave $K\pi$ system. It is instructive to compare
the \emph{K-matrix} against this more commonly employed isobar fit and we will
present and discuss the results for the two approaches. A third approach,
namely a partial-wave analysis of the $K \pi \pi$ system, will be considered
in a future publication.

\subsection{The isobar model}

The decay amplitude in this formalism is written as a coherent sum of
amplitudes corresponding to a constant term for the uniform direct three-body
decay and to different resonant channels:
\begin{equation}
\label{A_tot_isobar} \mathcal{M} =  a_0 e^{i\delta_0}+ \sum_j a_j~
e^{i\delta_j}~B(abc|r),
\end{equation}

where $a,b$ and $c$ label the final-state particles. Coefficient and phase of
the $K^*(892)$, our reference amplitude, are fixed to 1 and 0 respectively.

 $B(abc|r)=B(a,b|r)S(a,c)$ where
$B(a,b|r)$ is the Breit--Wigner function
\begin{equation}
\label{ff}
 B(a,b|r) =\frac{F_DF_r}{M^2_r -M^2_{ab}-i\Gamma M_r},
\end{equation}

and $S(a,c)=1$ for a spin-0 resonance, $S(a,c) =-2{\bf a} \cdot {\bf c}$ for a
spin-1 resonance and $S(a,c)\,=\,2\left(3 ({\bf a}\cdot {\bf c})^2 - |a|^2
|c|^2\right)$ for a spin-2 state. The ${\bf a}$ and ${\bf c}$ are the
three-momenta of particles $a$ and $c$ measured in the $ab$ rest frame.
The momentum-dependent form factors $F_D$ and $F_r$ represent the strong
coupling at each decay vertex, and are of the Blatt--Weisskopf form. For each
resonance of mass $M_r$ and spin $J$ we use a width

\begin{equation}
\label{gamma}
 \Gamma = \Gamma_r
\left(\frac{p}{p_r}\right)^{2J+1}\frac{M_r}{M_{ab}}\,
\frac{F^2_r(p)}{F^2_r(p_r)},
\end{equation}

where $p$ is the decay three-momentum in the resonance rest frame and the
subscript $r$ denotes the on-shell values. The order of particle label is
important for defining the phase convention; here the first particle is the
opposite-sign one, i.e., the kaon.
Each decay amplitude is then Bose-symmetrized with
respect to the exchange of the two identical pions.

\subsection{The K-matrix model}

The model of $S$-wave states requires particular care in order to account for
non-trivial dynamics generated by the presence of broad and overlapping
resonances: a real \emph{K-matrix} guarantees unitarity for two-body
interactions. An additional complication in the $K\pi$ system comes from the
presence in the $S$-wave of the two isospin states, $I=1/2$ and $I=3/2$.

The \emph{K-matrix} form we use as input describes the $S$-wave $K^-\pi^+ \to
K^-\pi^+$ scattering from the LASS experiment \cite{lass} for energy above 825
MeV and $K^-\pi^- \to K^- \pi^-$ scattering from Estabrooks \emph{et al.}
\cite{estabrook}. The \emph{K-matrix} form follows the extrapolation down to
$K\pi$ threshold for both $I=1/2$ and $I=3/2$ $S$-wave components by the
dispersive analysis by B\"uttiker \emph{et al.} \cite{butt}, consistent with
Chiral Perturbation Theory \cite{cpt}. The complete form is given below in
Eqs.~(\ref{eqn_T_tot}-\ref{eqn_K32}) with the parameters listed in
Table~\ref{tab_kmat12_value} \cite{mike_priv_com}.

Although only the $I=1/2$ is dominated by resonances, both isospin components
are involved in the decay of the $D^+$ meson into $K^-\pi^+\pi^+$. A model for
the decay amplitudes of the two isospin states can be constructed from the 2
$\times$ 2 \emph{K-matrix} describing the $I=1/2$ $S$-wave scattering in
$(K\pi)_1$ and $(K\eta')_2$ (with the subscripts 1 and 2, respectively,
labelling these two channels), and
 the single-channel \emph{K-matrix} describing the $I=3/2$ $K^-\pi^+
\to K^-\pi^+$ scattering.

The total $D$ decay amplitude in Eq.~(\ref{A_tot_isobar}) can be written as

\begin{equation}
\mathcal{M}= {(F_{1/2})}_1(s) + F_{3/2}(s) + \sum_j a_j~
e^{i\delta_j}~B(abc|r), \label{A_tot_kmat}
\end{equation}

where $s=M^2(K\pi)$, ${(F_{1/2})}_1$ and $F_{3/2}$ represent the $I=1/2$ and
$I=3/2$ decay amplitudes in the $K\pi$ channel, $j$ runs over vector and
spin-2 tensor resonances~\footnote{Higher spin resonances have been tried in
the fit with both formalisms but found to be statistically insignificant.},
and $~B(abc|r)$ are Breit--Wigner forms as in Eq.~(\ref{A_tot_isobar}) and
Eq.~(\ref{ff}). The $J>0$ resonances should, in principle, be treated in the
same \emph{K-matrix} formalism. However, the contribution from the vector wave
comes mainly from the $K^*(892)$ state, which is  well separated from the
higher mass $K^*(1410)$ and $K^*(1680)$, and the contribution from the spin-2
wave comes from $K_2^*(1430)$ alone. Their contributions are limited to small
percentages, and, as a first approximation, they can be reasonably described
by a simple sum of Breit--Wigners. More precise results would require a better
treatment of the overlapping $K^*(1410)$ and $K^*(1680)$ resonances as well.
$F_{1/2}$ is actually a vector consisting of two components: the first
accounting for the description of the $K\pi$ channel, the second of the
$K\eta'$ channel: in fitting $D^+ \to K^-\pi^+\pi^+$ we need, of course, the
${(F_{1/2})}_1$ element. Its form is

\begin{equation}
(F_{1/2})_1= (I-iK_{1/2}\rho)_{1j}^{-1}(P_{1/2})_j,
 \label{F_12}
\end{equation}

where $I$ is the identity matrix, $K_{1/2}$ is the \emph{K-matrix} for the
$I=1/2$ $S$-wave scattering in $K\pi$ and $K\eta'$, $\rho$ is the
corresponding phase-space matrix for the two channels \cite{chung} and
$(P_{1/2})_j$ is the production vector in the channel $j$. In this model
\cite{aitch}, the production process can be viewed as consisting of an initial
preparation of states, described by the \emph{P-vector}, which  then
propagates according to \mbox{$(I-iK\rho)^{-1}$} into the final one.

The form for $F_{3/2}$ is
\begin{equation}
F_{3/2}= (I-iK_{3/2}\rho)^{-1}P_{3/2},
 \label{F_32}
\end{equation}

where $K_{3/2}$ is the single-channel scalar function describing the $I=3/2$
\, \mbox{$K^-\pi^+ \to K^-\pi^+$} scattering, and $P_{3/2}$ is the production
function into $K\pi$.

The $K_I$ matrix for the isospin $I$ state is derived by fitting scattering
data via the corresponding $\mathcal{T}_I$ matrix defined as

\begin{equation}
\label{eqn_T_tot}
 \mbox{$\mathcal{T}_I=(I-iK_I\rho)^{-1}K_I$}.
\end{equation}

The ($\mathcal{T}_{1/2})_{11}$ and $\mathcal{T}_{3/2}$ functions are composed
to fit the $K^-\pi^+ \to K^-\pi^+$ $S$-wave through the Clebsch--Gordan
coefficients to give \mbox{$\mathcal{T}_{11} = (\mathcal{T}_{1/2})_{11} +
\mathcal{T}_{3/2}/2$}.

\begin{figure}[h]
  \centering
  \includegraphics[width=0.49\textwidth]{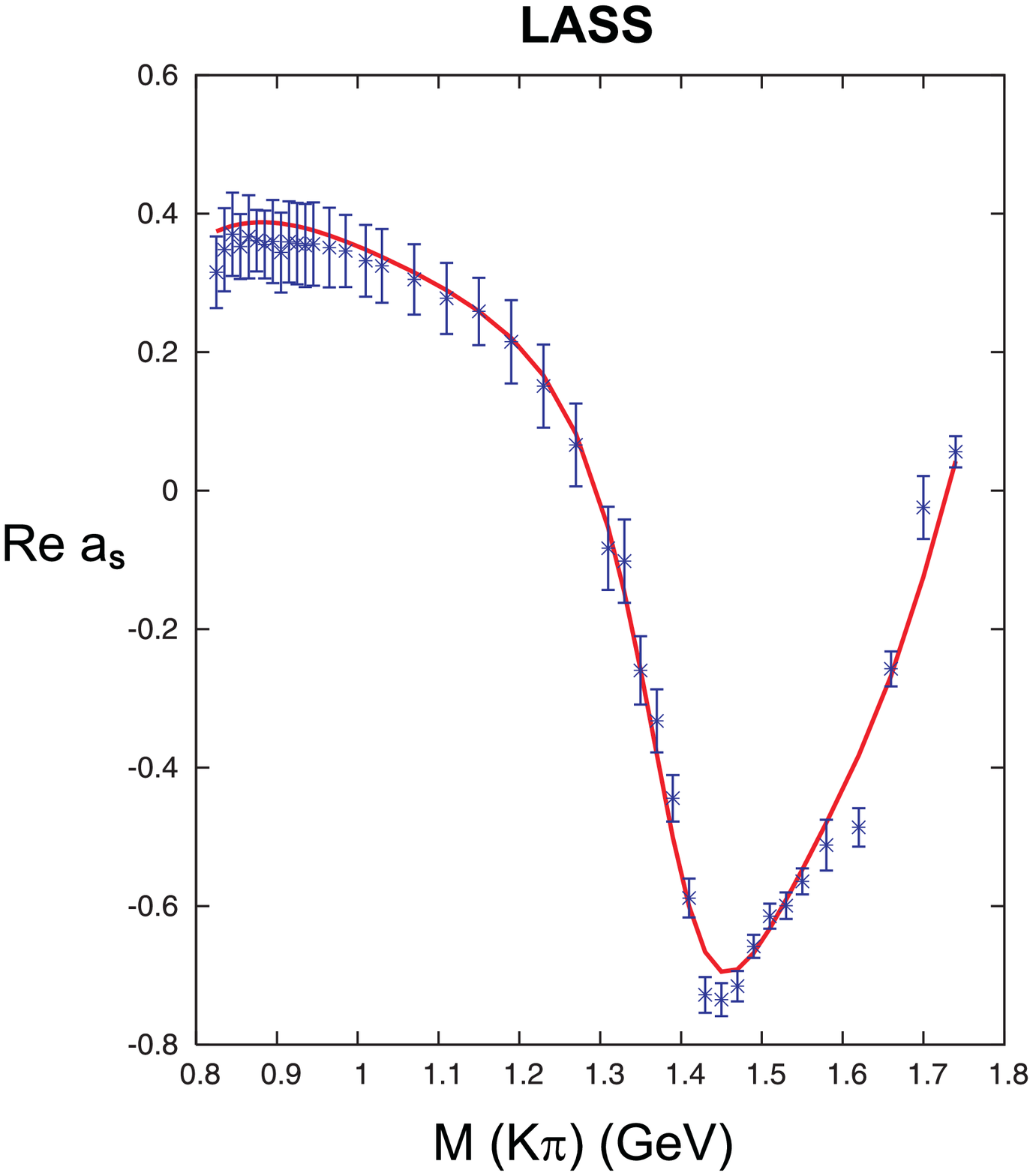}\hfil
  \includegraphics[width=0.49\textwidth]{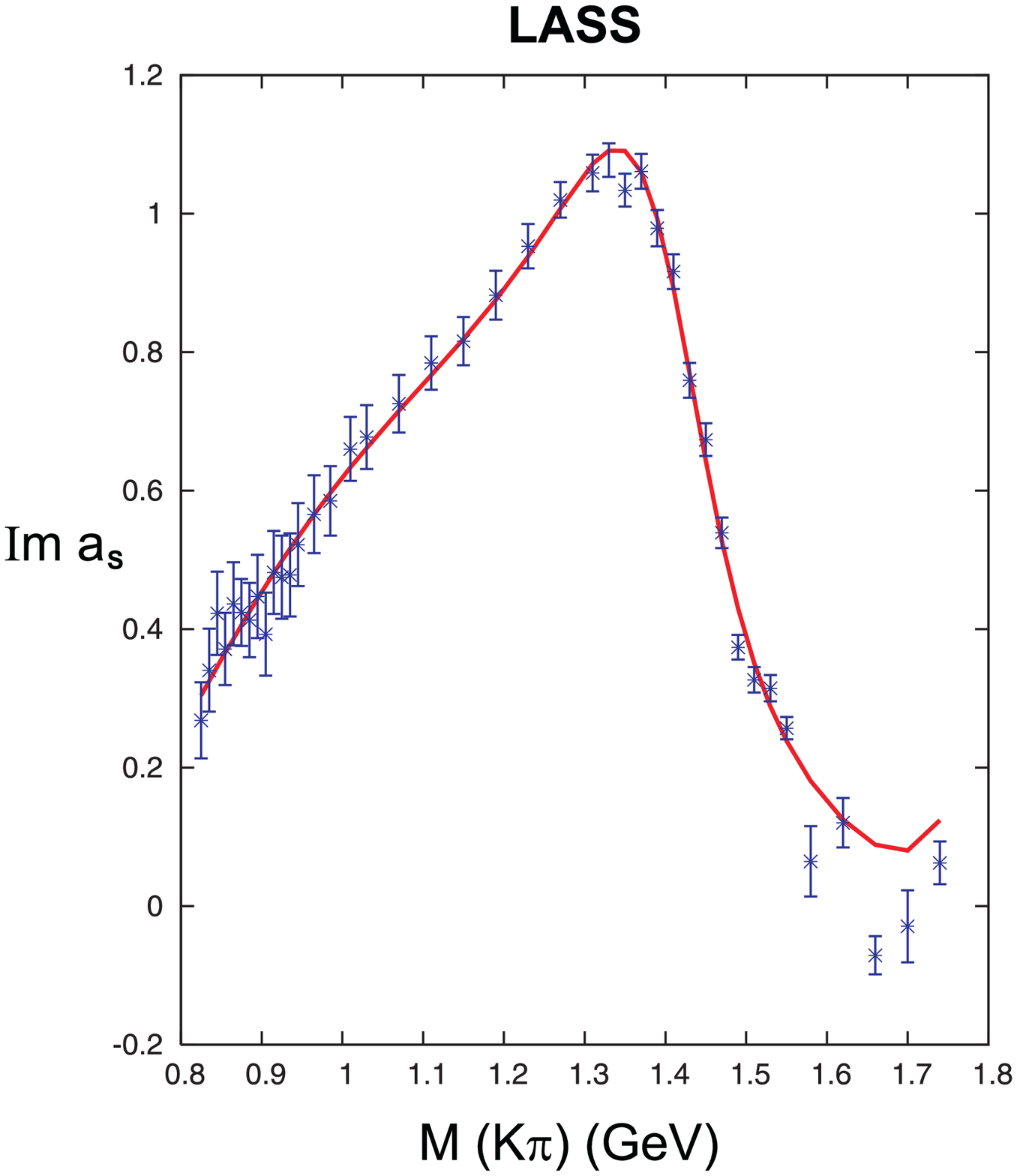}
  \caption{Real and imaginary $K^-\pi^+ \to K^-\pi^+$ amplitudes from the LASS experiment and their \emph{K-matrix} fit results. }
  \label{Lass_data}
\end{figure}

Fitting of the real and imaginary parts of the $K^-\pi^+ \to K^-\pi^+$ LASS
amplitude, shown in Fig.~\ref{Lass_data}, and using the predictions of Chiral
Perturbation Theory to continue this to threshold, gives the \emph{K-matrix}
parameters in Table~\ref{tab_kmat12_value}.

The $I=1/2$ \emph{K-matrix} is a one-pole, two-channel matrix whose elements
are given in Eq.~(\ref{eqn_K12}).

{\setlength\arraycolsep{2pt}
\begin{eqnarray}
 \label{eqn_K12}
 K_{11} &=& \left(\frac{s-s_{0\frac{1}{2}}}{s_{\text{norm}}} \right)  \left(\frac{g_1 \cdot g_1}{s_1-s}+
C_{110} + C_{111}\tilde {s} + C_{112}\tilde {s}^2 \right) \nonumber \\
K_{22} &=&  \left(\frac{s-s_{0\frac{1}{2}}}{s_{\text{norm}}} \right)
\left(\frac{g_2 \cdot g_2}{s_1-s}+ C_{220} + C_{221}\tilde {s} + C_{222}\tilde {s}^2\right) \nonumber\\
K_{12} &=& \left(\frac{s-s_{0\frac{1}{2}}}{s_{\text{norm}}} \right)
\left(\frac{g_1 \cdot g_2}{s_1-s} + C_{120} + C_{121}\tilde {s} +
C_{122}\tilde {s}^2\right),
\end{eqnarray}

where the factor of $s_{norm} =  m_K^2+ m_{\pi}^2$ is conveniently introduced
to make the individual terms in the above expression dimensionless. $g_1$ and
$g_2$ are the real couplings of the $s_1$ pole to the first and the second
channel respectively. \mbox{$s_{0\frac{1}{2}}=0.23$ GeV$^2$} is the Adler zero
position in the $I=1/2$ ChPT elastic scattering amplitude \footnote{ Chiral
symmetry breaking demands an Adler zero in the elastic $S$-wave amplitudes in
the unphysical region. ChPT at next-to-leading order fixes these positions
$s_{0I}$\cite{butt,cpt}.}. $C_{11i}$, $C_{22i}$ and $C_{12i}$ for $i=0,1,2$
are the three coefficients of a second order polynomial for the diagonal and
off-diagonal elements of the symmetric \emph{K-matrix}. Polynomials are
expanded around $\tilde s = s/s_{norm} -1$. This form generates an
\emph{S-matrix} pole, which is conventionally quoted in the complex energy
plane as $E=M -i\Gamma/2=1.408 -{\it i} 0.110$ GeV. Any more distant pole than
$K_0^*(1430)$ is not reliably determined as this simple \emph{K-matrix}
expression does not have the required analyticity properties. Nevertheless, it
is an accurate description for real values of the energy where scattering
takes place. Numerical values  of the terms in Eq.~(\ref{eqn_K12}) are
reported in Table~\ref{tab_kmat12_value}.

\begin{table}[!htb]
 \begin{center}
 \begin{tabular}{ccccc}
 \hline
 \emph{pole} (GeV$^2$) & \emph{coupling} (GeV) & $C_{11i}$ & $C_{12i}$ & $C_{22i}$  \\
 \hline
 \hline
 $ s_1=1.7919  $     &         &    &      & \\
          &   $ g_1= 0.31072$  &    &      & \\
          &   $ g_2=-0.02323$  &    &      & \\
          &          &        $C_{110}= 0.79299$ & $C_{120}= 0.15040$    &  $C_{220}=0.17054$     \\
          &          &        $C_{111}=-0.15099$ & $C_{121}= -0.038266$  &  $C_{221}=-0.0219$     \\
          &          &        $C_{112}= 0.00811$ & $C_{122}= 0.0022596$  &  $C_{222}=0.00085655$  \\
 \hline
  \end{tabular}
  \\[1ex]
  \caption{Values of parameters for the $I=1/2$  \emph{K-matrix}.}
   \label{tab_kmat12_value}
 \end{center}
\end{table}

The $I=3/2$ \emph{K-matrix} is given in Eq.~(\ref{eqn_K32}). Its form is
derived from a simultaneous fit to LASS data \cite{lass} and to $K^-\pi^- \to
K^-\pi^-$ scattering data \cite{estabrook}. It is a non-resonant, one channel
scalar function.

\begin{equation}
\label{eqn_K32}
 K_{3/2} = \left(\frac{s-s_{0\frac{3}{2}}}{s_{\text{norm}}} \right)
\left( D_{110} + D_{111} \tilde {s} + D_{112} \tilde {s}^2 \right).
\end{equation}

In Eq.~(\ref{eqn_K32}) $s_{0\frac{3}{2}}= 0.27$ GeV$^2$ is the Adler zero
position in the $I=3/2$ ChPT elastic scattering and the values of the
polynomial coefficients are \mbox{$D_{110} = -0.22147$}, \mbox{$D_{111}$ =
0.026637}, and \mbox{$D_{112} = -0.00092057$} \cite{mike_priv_com}.

When moving from scattering processes to $D$-decays, the production
\emph{P-vector} has to be introduced. While the \emph{K-matrix} is real,
\emph{P-vectors} are in general complex reflecting the fact that the initial
coupling $D^+\to (K^- \pi^+)\pi^+_{spectator}$ need not be real. The
\emph{P-vector} has to have the same poles as the \emph{K-matrix}, so that
these cancel in the physical decay amplitude. Their functional forms are:

\begin{equation}
(P_{1/2})_1= \frac{\beta g_1 e^{i\theta}} {s_1-s}
 + ( c_{10} + c_{11}\widehat{s}  + c_{12}
{\widehat{s}}^2 ) e^{i\gamma_1} \label{P_121}
\end{equation}

\begin{equation}
(P_{1/2})_2= \frac{\beta g_2 e^{i\theta}} {s_1-s} + (c_{20} + c_{21}\widehat
{s} +c_{22}{\widehat{s}}^2)e^{i\gamma_2} \label{P_122}
\end{equation}

\begin{equation}
P_{3/2}= (c_{30} + c_{31}\widehat{s} +c_{32}{\widehat{s}}^2)e^{i\gamma_3}
 \label{P_32}.
\end{equation}

$\beta e^{i\theta} $ is the complex coupling to the pole in the `initial'
production process, $g_1$ and $g_2$ are the couplings as given by
Table~\ref{tab_kmat12_value}. The $K\pi$ mass squared \mbox{$s_c = 2$ GeV$^2$}
corresponds to the centre of the Dalitz plot. It is convenient to choose this
as the value of $s$  about which the polynomials of
Eqs.~(\ref{P_121}-\ref{P_32}) are expanded, by defining $\widehat s = s -s_c$.
The polynomial terms in each channel are chosen to have a common phase
$\gamma_i$ to limit the number of free parameters in the fit and avoid
uncontrolled interference among the physical background terms. Thus the
coefficients of the second order polynomial, $c_{ij}$, are real. Coefficients
and phases of the \emph{P-vectors}, except $g_1$ and $g_2$, are the only free
parameters of the fit determining the scalar components.

\subsection{The likelihood function and fitting procedure}

In analogy with our previous works
we perform a fit to the $D^+$ Dalitz plot with an unbinned likelihood
function, $\mathcal{L}$, consisting of signal and background probability
density. The signal probability density is corrected for geometrical
acceptance and reconstruction efficiency. The shape of the background in the
signal region is fixed to that derived from a fit to the Dalitz plot of the
mass sidebands. It is parametrized through an incoherent sum of a polynomial
function plus resonant Breit--Wigner-like components. The signal fraction is
estimated by a fit to the $K^-\pi^+\pi^+$ mass spectrum. Checks for fitting
procedure are made using Monte Carlo techniques and all biases are found to be
small compared to the statistical errors. The systematic errors on our results
include \emph{split-sample} and \emph{fit-variant} components
\cite{K_matrix,Laura_dalitz}.
The sample is split in three different ways: low and high $D$ momenta, $D^+$
and $D^-$, and according to the two main running periods of the experiment.\\
The \emph{fit-variant} systematics is evaluated by varying the background
parametrization used to fit the sideband Dalitz plot, moving the sideband
regions, and letting the background freely float within the error returned by
the sideband fit.

\subsection{Isobar model results}

We allow for the possibility of contributions from all known well-established
($K^- \pi^+$) resonances\cite{pdg}. In addition, a constant amplitude accounts
for the direct decay of the $D$ meson into a non-resonant three-body final
states. The fit parameters are amplitude coefficients $a_i$ and phases
$\delta_i$ of Eq.~(\ref{A_tot_isobar}).
%
%
Contributions are removed if their amplitude coefficients, $a_i$ have less
than $2\,\sigma$ significance and the fit confidence level increases. The fit
confidence levels (C.L.) are evaluated with a $\chi^2$ estimator over the
Dalitz plot with bin size adaptively chosen to maintain a minimum number of
events in each bin. A fit consisting solely of well-established $K^-\pi^+$
resonances with masses and widths as in the PDG \cite{pdg} results in a very
poor solution, with an adaptive bin $\chi^2$/d.o.f of more than 3, and with a
very high level of the non-resonant component, about 90\% of the total,
atypically high for charm decays.

Following a previous work \cite{E791_kappa} a low mass $K^-\pi^+$ resonance
$\kappa$ is introduced with mass and width freely floating in the fit.  The
insertion of this state, whose mass and width are fitted to be \mbox{883 $\pm$
13 MeV/c$^2$} and \mbox{355 $\pm$ 13 MeV/$c^2$} respectively, returns a
$\chi^2$/d.o.f of more than 2. Only a simultaneous redefinition of the
Breit--Wigner parameters for the higher-mass scalar $K^*_0(1430)$ and the
inclusion of $\kappa$ gives an acceptable fit with a $\chi^2$/d.o.f of 1.17,
corresponding to a 6.8\% C.L. The mass and width of this effective
$K^*_0(1430)$ are \mbox{1461 $\pm$ 4 MeV/c$^2$} and \mbox{177 $\pm$ 8
MeV/c$^2$}, respectively; they can be compared to the PDG values of mass
\mbox{1412 $\pm$ 6 MeV/$c^2$} and width \mbox{294 $\pm$ 23 MeV/$c^2$}, and to
the \emph{K-matrix} fit values of mass 1408 MeV/$c^2$ and width 220 MeV/$c^2$.
The parameters for the $\kappa$ in a fit with a free $K^*_0(1430)$ are mass
\mbox{856 $\pm$ 17 MeV/$c^2$} and width \mbox{464 $\pm$ 28 MeV/$c^2$}. In
Table~\ref{tab_isobar_focus}, fit fractions\footnote{ The quoted fit fractions
are defined as the ratio between the intensity for a single amplitude
integrated over the Dalitz plot and that of the total amplitude with all the
modes and interferences present.}, phases and coefficients of the various
amplitudes from the fit are reported, along with masses and widths for
$\kappa$ and $K^*_0(1430)$.
\begin{table}[!htb]
\begin{center}
 \begin{tabular}{|cccc|}
 \hline
 \emph{channel} & \emph{fit fraction (\%)} & \emph{phase $\delta_i$ (deg)}
 & \emph{coefficient} \\
 \hline
 \hline
 $non-resonant$      &  $29.7 \pm 4.5$                    & $325  \pm 4 $          & $1.47   \pm 0.11$ \\
                     &  $ \pm \ 1.5 \pm 2.1$ (see text)      & $\pm \ 2 \pm 1.2$        & $\pm \ 0.06 \pm 0.06$      \\
 $K^*(892)\pi^+$     &  $13.7 \pm 0.9  $     & $0$  (fixed)                & $1$(fixed)                      \\
                     &  $ \pm \ 0.6 \pm 0.3$      &                             & \\
 $K^*(1410)\pi^+$    &  $ 0.2  \pm 0.1 $      & $ 350 \pm 34   $          & $ 0.12  \pm 0.03 $     \\
                     &  $ \pm  \ 0.1 \pm 0.04 $ & $\pm \ 17 \pm 15 $          & $\pm \ 0.003 \pm   0.01$ \\
 $K^*(1680)\pi^+$    &  $1.8  \pm 0.4 $      & $ 3   \pm  7  $           & $ 0.36  \pm 0.04 $      \\
                     &  $ \pm \ 0.2 \pm 0.3  $  & $\pm \ 4 \pm 8 $            & $ \pm \ 0.02 \pm     0.03$ \\
 $K^*_2(1430)\pi^+$  &  $ 0.4  \pm 0.05 $      & $ 319 \pm 8      $        & $ 0.17  \pm 0.01 $     \\
                     &  $ \pm \ 0.04 \pm 0.03$   & $ \pm \ 2 \pm 2$            & $\pm \ 0.01 \pm 0.01 $ \\
 $K^*_0(1430)\pi^+$  &  $17.5 \pm 1.5  $     & $  36 \pm  5    $         & $ 1.13  \pm 0.05 $     \\
                     &  $ \pm \ 0.8 \pm 0.4 $     & $  \pm \ 2 \pm 1.2 $        & $ \pm \ 0.01 \pm  0.02$ \\
 $\kappa\pi^+$       &  $22.4 \pm 3.7 $       & $199  \pm 6    $     & $1.28   \pm 0.10 $     \\
                     & $  \pm \ 1.2 \pm 1.5 $ (see text)   & $\pm \ 1  \pm 5  $          & $\pm \ 0.015 \pm 0.04  $ \\
\hline \hline
& mass (MeV/$c^2$) & width (MeV/$c^2$) &   \\
 $K^*_0(1430)$ &
  $1461 \pm 4 \pm 2 \pm 0.5$   & $ 177\pm  8 \pm 3 \pm 1.5 $ & \\
$\kappa$
&$856 \pm 17 \pm 5 \pm 12$ & $464  \pm 28 \pm 6 \pm 21$ & \\
 \hline
 \hline
 \end{tabular}
 \\[1ex]
 \caption{Fit fractions, phases, and coefficients from the isobar fit
 to the FOCUS $D^+ \to K^-\pi^+\pi^+$ data. The first error is statistic, the second error
is systematic from the experiment, and the third error is systematic induced
by model input parameters for higher resonances.}
 \label{tab_isobar_focus}
 \end{center}
\end{table}
Coefficients refer to Bose-symmetrized, normalized amplitudes, both for
resonant and non-resonant states. Masses and widths can be compared with
previous determinations from E791 \cite{E791_kappa} and BES \cite{Bes_kappa}.
All the results are consistent within the errors.
%
Two systematic uncertainties are quoted in Table~\ref{tab_isobar_focus}\,: the
first includes our \emph{split sample} and \emph{fit variant} components of
Section 3.3, the second reflects uncertainties due to higher resonance
modeling. In particular, this analysis is performed with a fixed radius of 5
GeV$^{-1}$ and 1.5 GeV$^{-1}$ in the Blatt--Weisskopf $D$-meson and resonance
form factors, respectively \cite{E791_kappa}. To cover the large range of
uncertainties we found in literature on these parameters, we estimated the
systematic effect by varying them. Furthermore, for $J>0$ resonances we
assumed central values for masses and widths from the PDG \cite{pdg}. For
those parameters which are not accurately determined we allowed a $\pm 1\,
\sigma$ variation and estimated the systematic uncertainty by the
corresponding spread in the results. An analogous study is performed for the
\emph{K-matrix} fit and the corresponding uncertainties are quoted in
Table~\ref{tab_P_vector_nov06} and \ref{tab_high_spin}.
\begin{figure}[!htb]
  \centering
  \includegraphics[width=0.98\textwidth]{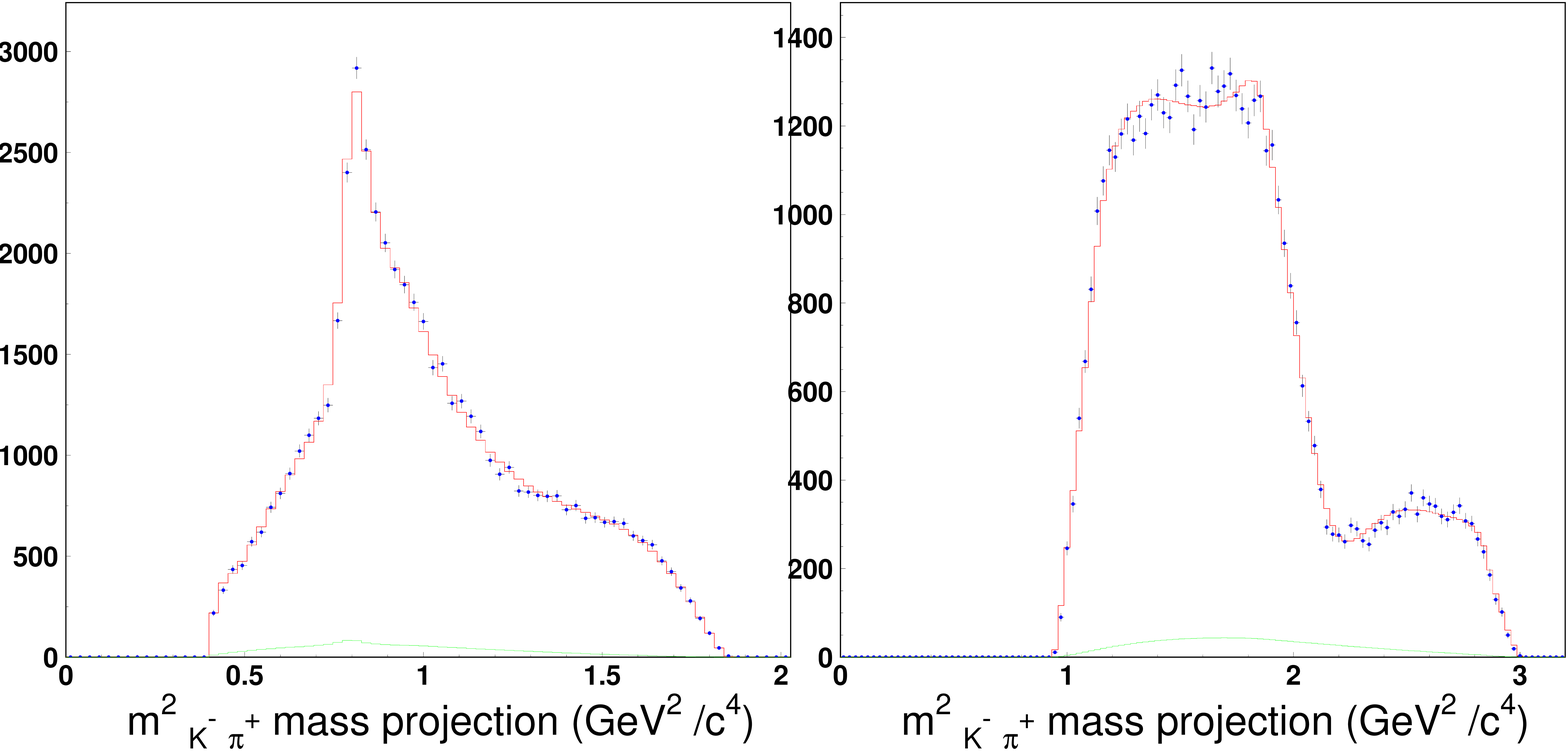}\\
  \includegraphics[width=0.49\textwidth]{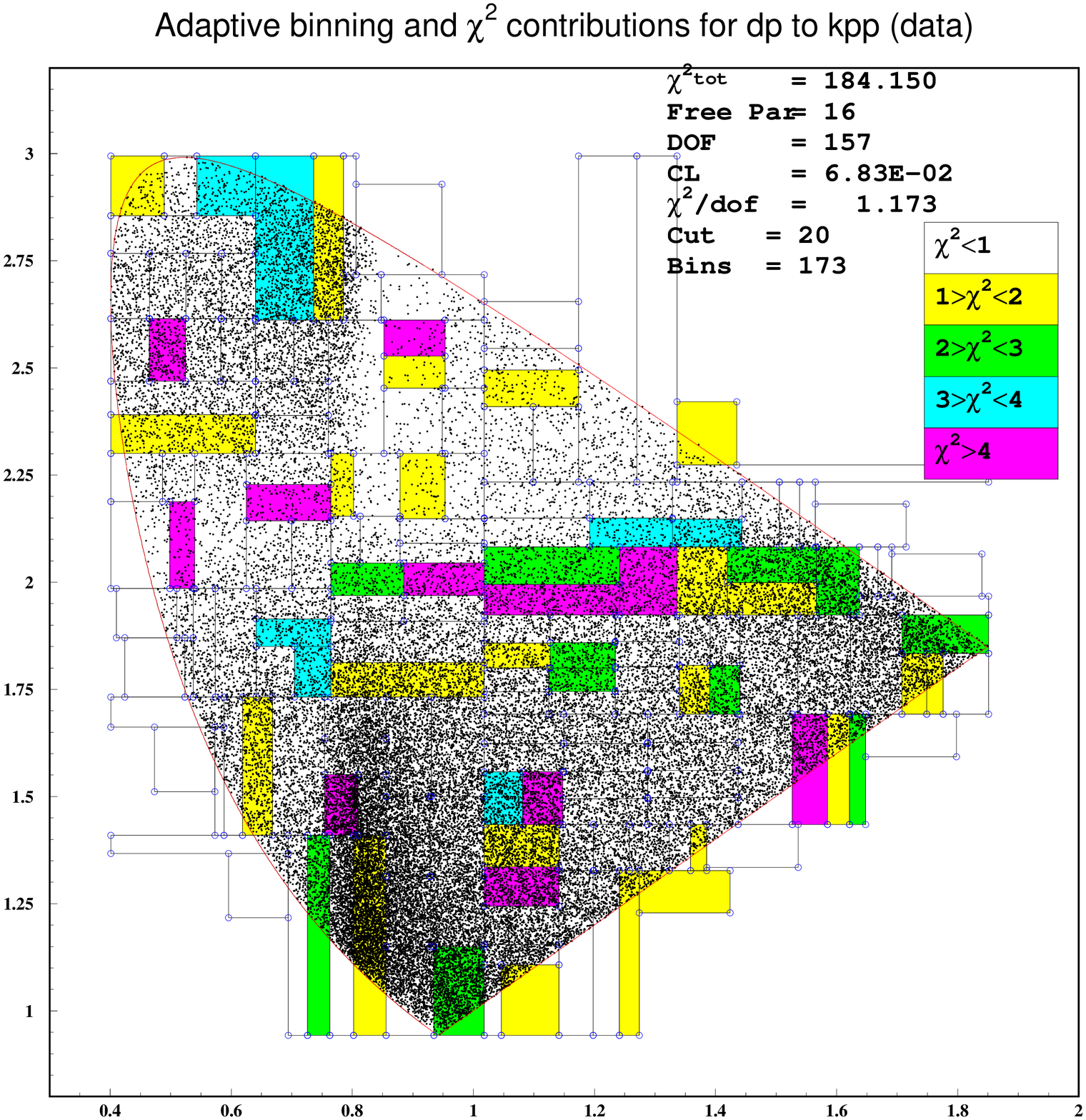}\hspace*{0.49\textwidth}
 \caption{ Top: Dalitz plot projections with our isobar fit superimposed.
  The background shape under the signal is also shown.
 Bottom: The adaptive binning scheme.}
  \label{fit_isobar_results}
\end{figure}
Systematics for the $\kappa$ and non-resonant components have a very large
dependence on whether or not Gaussian form factors, suggested in
\cite{ton_FF}, are included in the fit. Our systematic uncertainties do not
account for the effect introduced by adding form factor for the scalar
resonances. There is still some controversy on whether form factors are even
needed for these scalars \cite{bugg_FF}. We find that adding in the Gaussian
form factors does not improve our fit quality (3\% C.L versus 6.8\% C.L.). In
this case, our results become consistent, within the errors, with Model C in
\cite{E791_kappa}, and changes the $\kappa$ and non-resonant fit fractions to
\mbox{(40.7 $\pm$ 6.2) \%} and \mbox{(17.5 $\pm$ 4.1 )\%} respectively. Dalitz
plot projections with fit results and the corresponding adaptive-binning
scheme are shown in Fig.~\ref{fit_isobar_results}.
%
%
\subsection{K-matrix model results}

In the \emph{K-matrix} fit to the $D^+ \to K^- \pi^+ \pi^+$ decay the free
parameters are amplitudes and phases ($a_i$ and $\delta_i$) for vectors and
tensors, and the \emph{P-vector} parameters for scalar contributions. $K\pi$
scattering determines the parameters of the {\it K-matrix} elements and these
are fixed inputs to this $D$ decay analysis.
Table~\ref{tab_P_vector_nov06} reports our \emph{K-matrix} fit results. It
shows quadratic terms in $(P_{1/2})_1$ are significant in fitting data, while
in both  $(P_{1/2})_2$ and $P_{3/2}$ constants are sufficient.
\begin{table}[!htb]
 \begin{center}
 \begin{tabular}{|cc|}
 \hline
 \emph{coefficient} & \emph{phase (deg)}\\
 \hline
 \hline
 $\beta$      = $3.389  \pm  0.152 \pm 0.002  \pm 0.068   $     & $\theta=    286 \pm 4 \pm 0.3 \pm 3.0   $  \\
 $c_{10}$     = $1.655  \pm  0.156 \pm 0.010  \pm 0.101   $     & $\gamma_1 = 304 \pm 6 \pm 0.4 \pm 5.8$ \\
 $c_{11}$     = $0.780  \pm  0.096 \pm 0.003  \pm 0.090   $     &      \\
 $c_{12}$     = $-0.954 \pm 0.058  \pm  0.0015 \pm 0.025  $     &      \\
 $c_{20}$     = $17.182 \pm 1.036  \pm 0.023 \pm 0.362    $     & $ \gamma_2 = 126  \pm  3 \pm 0.1 \pm 1.2  $ \\
 $c_{30}$     = $0.734  \pm 0.080  \pm 0.005 \pm 0.030    $     & $ \gamma_3 = 211  \pm 10 \pm 0.7\pm 7.8 $ \\
 \hline
 \multicolumn{2}{|c|}{
\emph{Total $S$-wave fit fraction}  = $ 83.23 \pm 1.50 \pm 0.04 \pm 0.07$ \%}  \\
 \multicolumn{2}{|c|}{
\emph{Isospin 1/2 fraction }  = $207.25 \pm 25.45 \pm  1.81 \pm 12.23$ \% } \\
 \multicolumn{2}{|c|}{
\emph{Isospin 3/2 fraction }  = $40.50  \pm 9.63 \pm 0.55 \pm 3.15$ \% } \\
 \hline
 \end{tabular}
 \\[1ex]
 \caption{$S$-wave  parameters from the \emph{K-matrix} fit
 to the FOCUS $D^+ \to K^-\pi^+\pi^+$ data.  The first error is statistic, the second error is systematic from the experiment,
 and the third is systematic induced by model input parameters for higher resonances.
 Coefficients are for the unnormalized $S$-wave.}
 \label{tab_P_vector_nov06}
 \end{center}
\end{table}

The $J>0$ states required by the fit are listed in Table~\ref{tab_high_spin}.

\begin{table}[!htb]
 \begin{center}
 \begin{tabular}{|cccc|}
 \hline
 \emph{component} & \emph{fit fraction (\%)} & \emph{phase $\delta_j$ (deg)}
 & \emph{coefficient} \\
 \hline
 \hline
 $K^*(892)\pi^+$     &  $13.61 \pm  0.98 $       &    0 (fixed)                          & 1 (fixed)          \\
                     &  $ \pm \ 0.01 \pm 0.30 $  &                                       &                    \\

 $K^*(1680)\pi^+$    &  $ 1.90 \pm  0.63  $      &  $1   \pm  7  $       & $0.373 \pm 0.067 $ \\
                     &  $ \pm \ 0.009 \pm 0.43$  &  $ \pm \ 0.1 \pm 6 $  & $ \pm \ 0.009 \pm 0.047$ \\
 $K^*_2(1430)\pi^+$  &  $ 0.39 \pm  0.09  $      &  $296 \pm 7   $       & $0.169 \pm 0.017$ \\
                     &  $\pm \ 0.004 \pm 0.05 $  & $\pm\  0.3 \pm 1 $    & $ \pm \ 0.010 \pm 0.012 $ \\
 $K^*(1410)\pi^+$    &  $ 0.48 \pm  0.21  $      &  $293 \pm 17  $       & $0.188 \pm 0.041 $ \\
                     &  $ \pm \ 0.012 \pm 0.17$  & $\pm \ 0.4 \pm 7 $    & $ \pm \ 0.002 \pm 0.030$ \\
 \hline
 \hline
 \end{tabular}
 \\[1ex]
 \caption{Fit fractions, phases, and coefficients for the $J>0$ components from the \emph{K-matrix} fit
 to the FOCUS  $D^+ \to K^-\pi^+\pi^+$ data.  The first error is statistic, the
 second error is systematic from the experiment, and the
 third error is systematic induced by model input parameters for higher resonances. }
 \label{tab_high_spin}
 \end{center}
\end{table}
The $S$-wave component accounts for the dominant portion of the decay $(83.23
\pm 1.50) \%$, and is the sum of the $I=1/2$ and $I=3/2$ components, which,
separately, account for the (207 $\pm$ 24)\% and (40 $\pm$ 9)\% of the decay,
with $-164$ \% from their interference. The large amount of interference
between the  $I=3/2$ component and the $I=1/2$ component underscores its
importance in our \emph{K-matrix} fit.  Because there are no $I=3/2$
resonances, the $I= 3/2$ $S$-wave component has at most a slowly varying phase
and amplitude and since we find little variation in the $F_{3/2}$-vector phase
as well, the $I=3/2$ piece essentially plays a comparable role to the $\sim$
30\% non-resonant component present in the isobar fit summarized by Table 2. A
significant fraction, $13.61\pm 0.98$\%, comes, as expected, from $K^*(892)$;
smaller contributions come from two vectors $K^*(1410)$ and $K^*(1680)$ and
from the tensor $K_2^*(1430)$.
%
%
It is conventional to quote fit fractions for each component and this is what
we do. Care should be taken in interpreting some of these since strong
interference can occur. This is particularly apparent between contributions in
the same-spin partial wave. While the total $S$-wave fraction is a sensitive
measure of its contribution to the Dalitz plot, the separate fit fractions for
$I=1/2$ and $I=3/2$ must be treated with care. The broad $I=1/2$ $S$-wave
component inevitably interferes strongly with the slowly varying $I=3/2$
$S$-wave, as seen for instance in \cite{mike_laura}.
Fit results on the projections and the adaptive binning scheme are shown in
Fig.~\ref{fit_kmatrix_nov06}.

The fit $\chi^2$/d.o.f is 1.27 corresponding to a confidence level of 1.2\%.
If the $I=3/2$ component is removed from the fit, the $\chi^2$/d.o.f worsens
to 1.54, corresponding to a confidence level of $10^{-5}$.
\begin{figure}[!htb]
 \centering
 \includegraphics[width=0.98\textwidth]{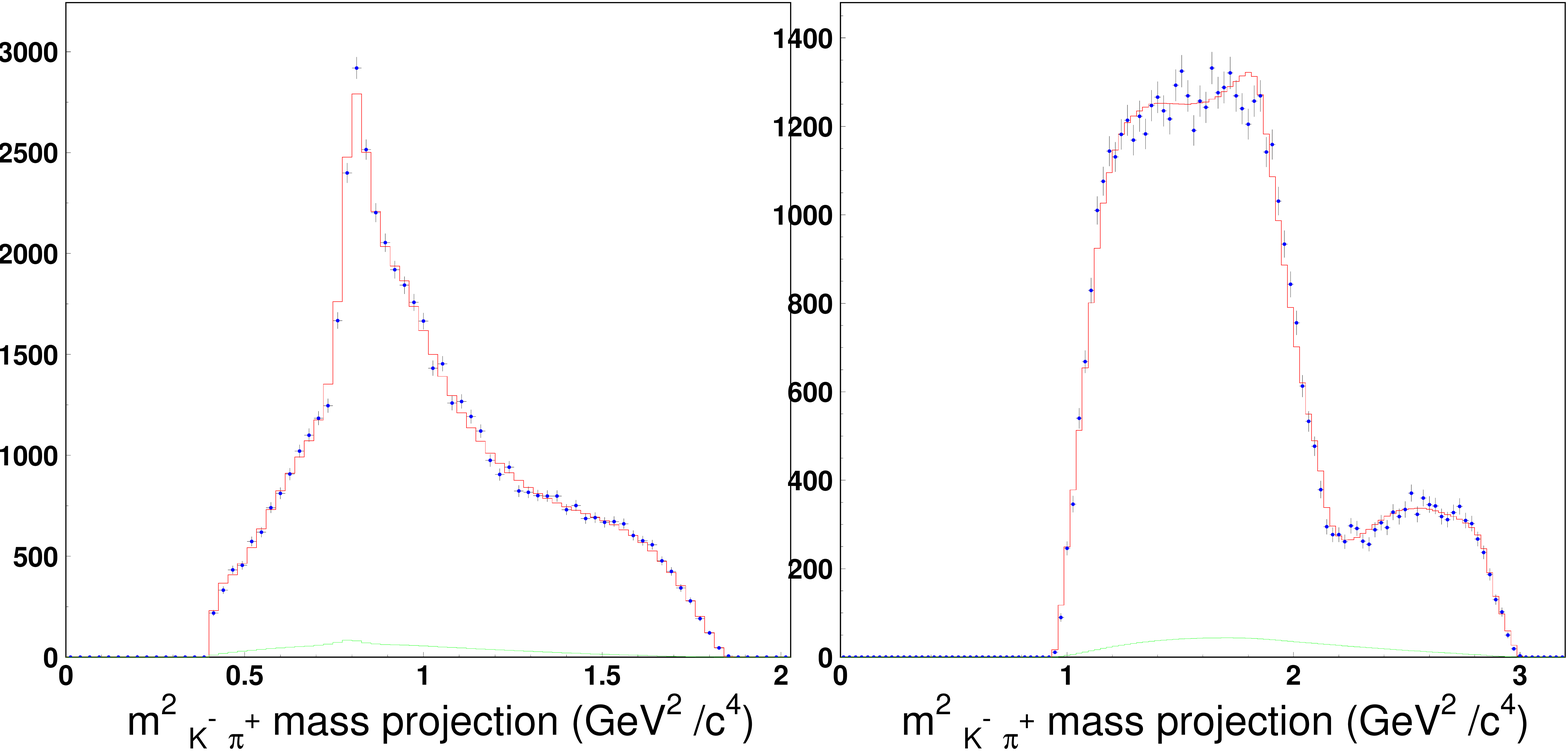}
 \includegraphics[width=0.49\textwidth]{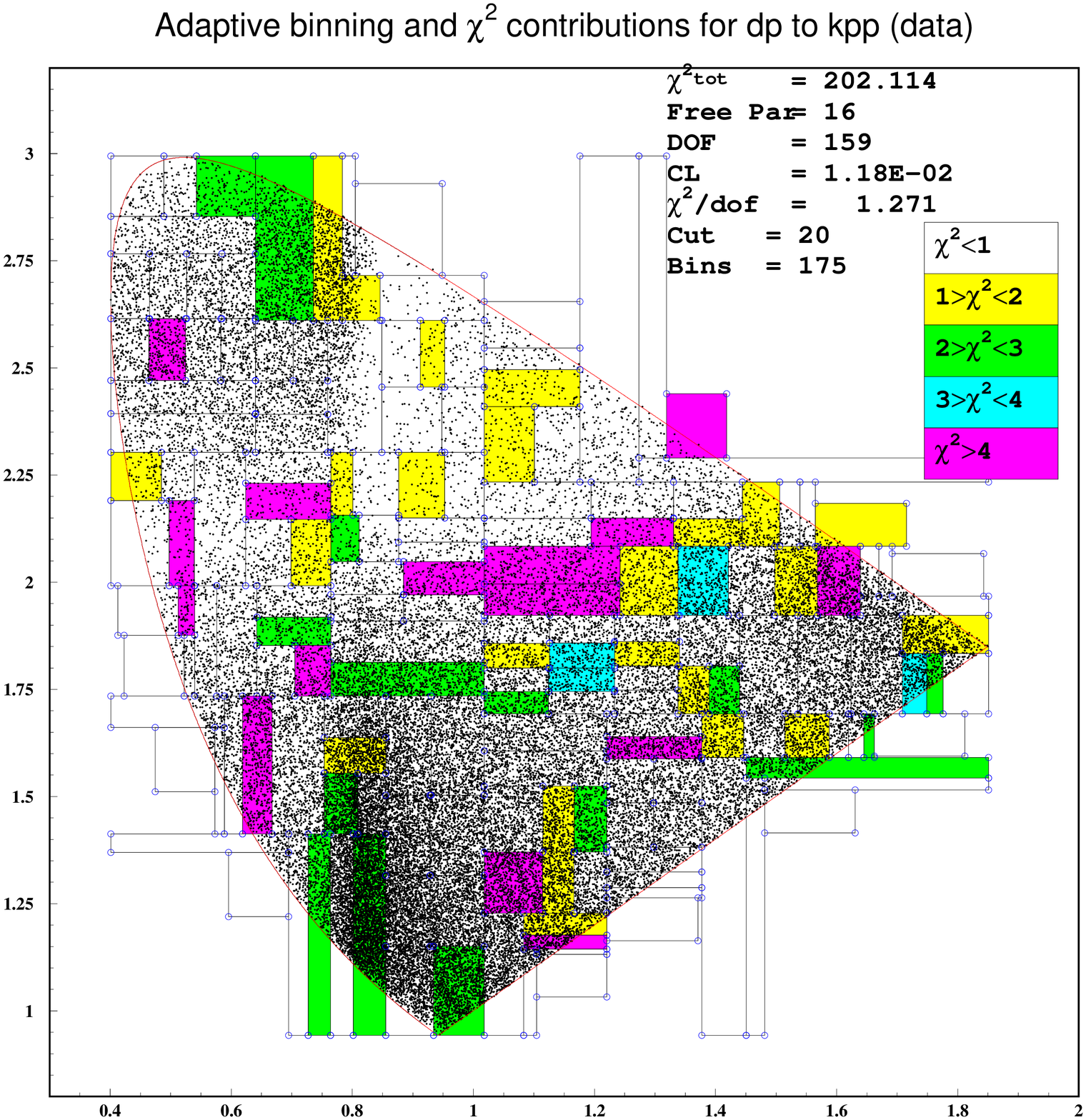}\hspace*{0.49\textwidth}
  \caption{Top: The Dalitz plot projections with the \emph{K-matrix} fit superimposed.
  The background shape under the signal is also shown.
 Bottom: The adaptive binning scheme.}
  \label{fit_kmatrix_nov06}
\end{figure}
%

\subsection{Comparison and discussion of the results}

The isobar fit represents a good effective description of the data, as
testified by the confidence level{}. However, simple Breit--Wigner forms have
been used for both the broad scalars $\kappa$ and $K^*_0(1430)$, each with
free mass and width with no reference to how these states appear in other
$K\pi$ interactions. Elastic scattering provides independent information about
these states, their shape and parameters and how they overlap. In particular,
scattering data from LASS show that the phase of the $S$-wave rises by no more
than $100^\circ$ from 825 to 1450 MeV, while the isobar fit with its simple
Breit--Wigner forms requires $\sim 180^\circ$ change in the resonant $I=1/2$
$S$-wave. In Fig.~\ref{kpi_kmatrix}a) and b) we compare the modulus and phase
of the total $S$-wave components from the isobar and \emph{K-matrix} fits.
They essentially agree, as expected, since they fit the same data. However,
the physics description differs considerably in the $I=1/2$ components.
In Fig.~\ref{kpi_kmatrix} c) and d) modulus and phase of the $I=1/2$ $S$-wave
component from our \emph{K-matrix} fit are shown.
\begin{figure}[!htb]
 \centering
\includegraphics[width=0.98\textwidth]{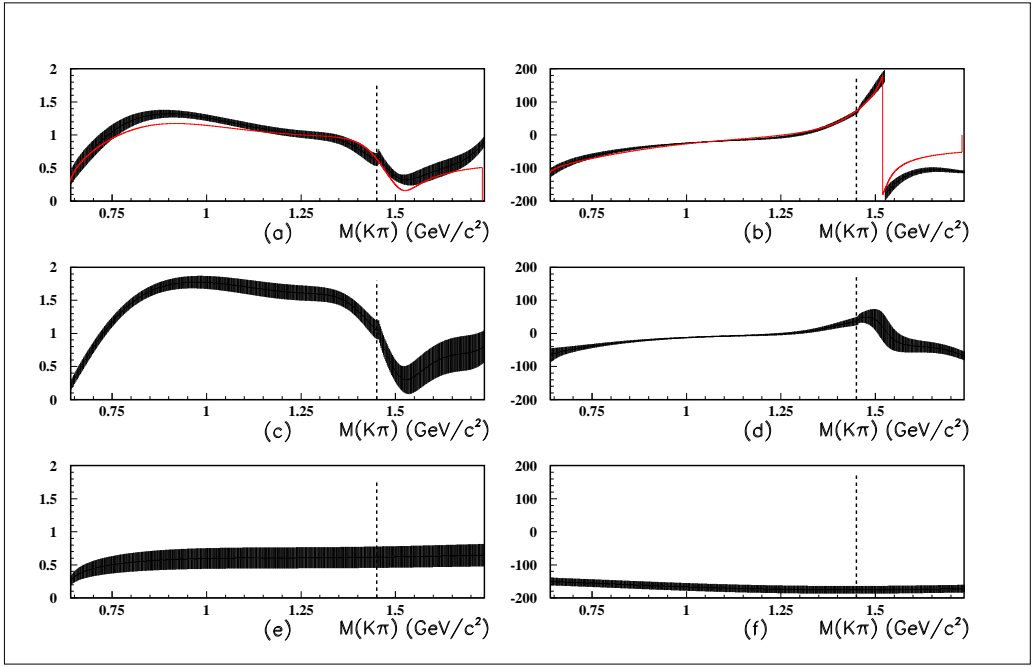}
\caption{Modulus and phase of the $S$-wave components resulting from the
 \emph{K-matrix} and isobar fit. Bands represent the $\pm 1 \,\, \sigma$
 statistical variation of the \emph{K-matrix} results
 for the total component [ a) and b)  plots], the $I=1/2$ component [ c) and d)
 plots)]
 and the the $I=3/2$ component [ e) and f)]  plots).
 Continuous lines in a) and b) represent the central value of the isobar
 fit. Vertical dashed line shows the location of the $K\eta'$ threshold.}
  \label{kpi_kmatrix}
\end{figure}
The motivation for the \emph{K-matrix} fit was to bring consistency between
the description of scattering and $D$-decay data. In such a formalism the
poles of the $S$-matrix are process independent, and on the real energy axis
the overlap of broad resonances is correctly described. The results of the
\emph{K-matrix} fit showed that such a consistent representation is equally
possible, the global fit quality being indeed good. However, it deteriorates
at higher $K\pi$ mass. This is not surprising since our {\it K-matrix}
treatment only includes two channels $K\pi$ and $K\eta'$. While we have
reliable information on the former channel, we have relatively poor
constraints on the latter. This means that as we consider $K\pi$ masses far
above $K\eta'$ threshold, these inadequacies in the description of the
$K\eta'$ channel become increasingly important. This is expected to become
worse as yet further inelastic channels open up. Consequently, improvements
could be made by using a number of $D$-decay chains with $K\pi$ final state
interactions and inputting all these in one combined analysis in which several
inelastic channels are included in the \emph{K-matrix} formalism. In the
present single $D^+\to K^-\pi^+\pi^+$ channel, adding further inelastic modes
would be just adding free unconstrained parameters for which there is little
justification. It is interesting to note that the adaptive binning scheme
shows that both the \emph{K-matrix} and the isobar fit are not able to
reproduce data well in the region at 2 GeV$^2$, in the vicinity of the
$K\eta'$ threshold. It is also the energy domain where higher spin states
live. Vector and tensor fit parameters in the two models are in very good
agreement: we do not exclude the possibility that a better treatment of these
amplitudes could improve the $\chi^2$. Some isolated spots of high $\chi^2$
could be caused by an imperfect modeling of the efficiency as they are in the
same regions in both fits. The FOCUS experiment has also studied the
$K^-\pi^+$ amplitudes in the $D^+ \to K^-K^+\pi^+$ decay. We note that the
moduli of Fig.~\ref{kpi_kmatrix}a) are very different than that presented in
Fig.~3a) of \cite{focus_kkp}. This underscores the fact that different decay
modes of the same charm state can often have quite different shapes for the
moduli of their amplitudes, reflecting the differing production dynamics
encoded in the \emph{P-vectors}.

\section{Conclusions}

The analysis of the $D^+\to K^-\pi^+\pi^+$ decay with the two models described
in this paper reveal quite different features. The isobar model with its
Breit--Wigner representation for all states requires both a $\kappa$ and a
 $K^*_0(1430)$ whose parameters are not what elastic scattering would require.
However, as already indicated,  such Breit--Wigner parameters are effective
rather than genuine pole positions.
 In contrast, the {\it K-matrix} fit has built in consistency with
 $K\pi$ scattering. Moreover, this agrees with the results of
 our FOCUS experiment in semi-leptonic $D$-decay
 \cite{jim_sl_1,jim_sl_2,massaferri}, where we have shown that the $S$-wave
 phase
agrees with that of the LASS experiment. The hypothesis of the two-body
dominance, which has already been tested in other charm meson decays, is
consistent with our results for the high-statistics $D^+ \to K^- \pi^+\pi^+$.
Previous analyses \cite{Brian} did not allow for the two different isospin
states, and compared the $I=1/2$ scattering phase with the global scalar
phase. Comparison with our global \emph{F-vector} phase is shown in the left
plot of Fig.~\ref{phase_tot_compa}\footnote{Phases determined from scattering
are absolute. Those from the $D^+ \to K^-\pi^+ \pi^+$ Dalitz analysis are
relative. We are free to raise the phases of Fig.~\ref{kpi_kmatrix} to be zero
at threshold.}.
\begin{figure}[h]
\centering
  \includegraphics[width=0.98\textwidth]{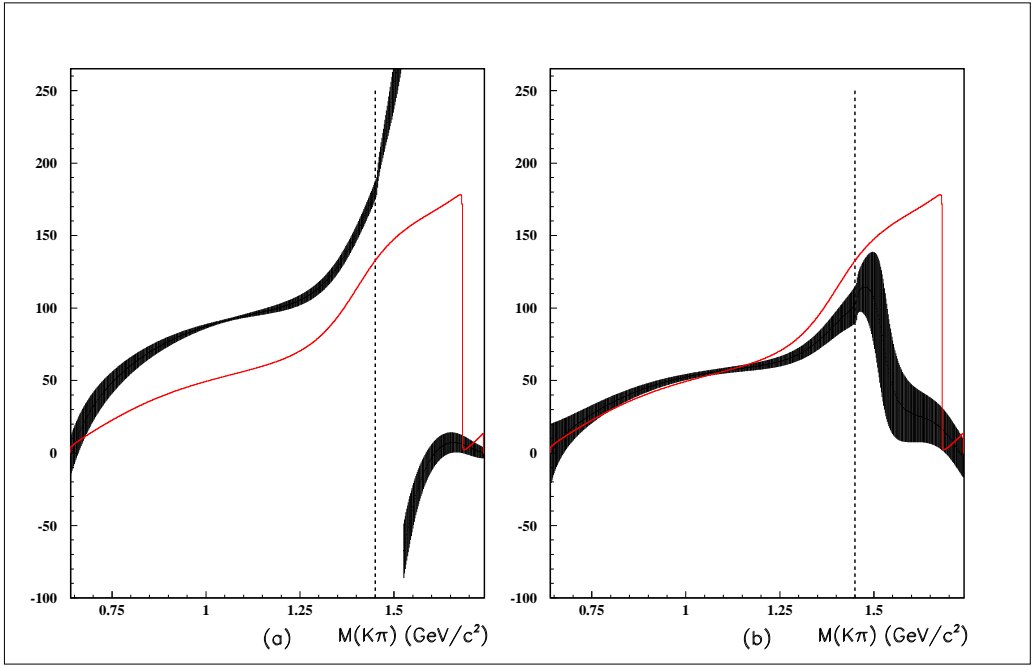}
  \caption{Comparison between the LASS $I=1/2$ phase + ChPT (continous line)
  and the \emph{F-vector} phases (with $\pm 1\,\sigma$ statical error bars); a)
   total \emph{F-vector} phase; b) $I=1/2$ \emph{F-vector} phase.
  Vertical dashed line shows the location of the $K\eta'$.}
  \label{phase_tot_compa}
\end{figure}
A feature of the \emph{K-matrix} amplitude analysis is that it allows an
indirect phase measurement of the separate isospin components: it is this
phase variation with isospin $I=1/2$ which should be compared with the same
$I=1/2$ LASS phase, extrapolated from 825 GeV down to threshold according to
Chiral Perturbation Theory. This is done in the right plot of
Fig.~\ref{phase_tot_compa}. As explained in Section 3.2, in this model
\cite{aitch} the \emph{P-vector} allows for a phase variation accounting for
the interaction with the third particle in the process of resonance formation.
It so happens that the Dalitz fit gives a nearly constant production phase.
The two phases in Fig.~\ref{phase_tot_compa}b) have the same behaviour up to
$\sim$ 1.1 GeV. However, approaching $K\eta'$ threshold, effects of
inelasticity and differing final state interactions start to appear.
The difference between the phases in Fig.~\ref{phase_tot_compa}a) is due to
the $I=3/2$ component.

These results are consistent with $K\pi$ scattering data, and consequently
with Watson's theorem predictions for two-body $K\pi$ interactions in the low
$K\pi$ mass region, up to $\sim$ 1.1 GeV, where elastic processes dominate.
This means that possible three-body interaction effects, not accounted for in
the \emph{K-matrix} parametrization, play a marginal role.

Our results for the total $S$-wave are in general agreement with those from
the E791 analysis, in which the $S$-wave modulus and phase were determined in
each $K\pi$ slice \cite{Brian}, \cite{Mike_china}.
Edera and Pennington \cite{mike_laura} were able to separate this total
$S$-wave into $I=1/2$ and $I=3/2$ components, making the strong assumption
that the equivalent of the \emph{P-vector} has a constant phase. Here we have
relaxed this assumption, and find a slightly different separation of $I=1/2$
and $3/2$ components, but the trend with energy and relative phases are
broadly consistent.

What does this analysis contribute to the discussion of the existence and
parameters of the $\kappa$?  We know from analysis \cite{cherry} of the LASS
data (which in $K^-\pi^+$ scattering only start at 825 MeV) there is no pole,
the $\kappa(900)$, in its energy range. However, below 800 MeV, deep in the
complex plane, there is very likely such a state. Its precise location
requires a more sophisticated analytic continuation onto the unphysical sheet
than the {\it K-matrix} representation provided here. This is because of the
need to approach close to the crossed channel cut, which is not correctly
represented for a robust analytic continuation. However, our {\it K-matrix}
representation fits along the real energy axis inputs on scattering data and
Chiral Perturbation Theory in close agreement with those used in the analysis
by Descotes-Genon and Moussallam \cite{Desco_Mussalla} that locates the
$\kappa$ with a mass of $(658 \pm 13)$ MeV and a width of $(557 \pm 24)$ MeV
by careful continuation. These pole parameters  are quite different from those
implied by the simple isobar fits presented here, and by E791 in
\cite{E791_kappa}. What we have shown is that whatever $\kappa$ is revealed by
our $D^+\to K^-\pi^+\pi^+$ results, it is the same as that found in scattering
data. Consequently, our analysis supports the conclusions of
\cite{Desco_Mussalla} and \cite{Zhou}.

We have seen that $D$-decay can teach us about $K\pi$ interaction much closer
to threshold than the older scattering results. This serves as a valuable
check from experiment \cite{sanmike} of the inputs to the analyses of
\cite{Desco_Mussalla} and \cite{Zhou} based largely on theoretical
considerations. Indeed, more complete experimental insight into the $K\pi$
interaction will be provided by the  full range of hadronic and semi-leptonic
$D$-decays to come from $B$-factories. Our results show that the dynamics of
the $K^-\pi^+\pi^+$ final state is dominated by two-body $K\pi$ interactions
up to 1.1 GeV as determined by scattering experiments.

\section{Acknowledgments}
We wish to acknowledge the assistance of the staffs of
Fermi National Accelerator Laboratory, the INFN of Italy, and the physics
departments of the collaborating institutions. This research was supported in
part by the U. S. National Science Fundation, the U. S. Department of Energy, the
Italian Istituto Nazionale di Fisica Nucleare and Ministero
dell'Universit\`a e della Ricerca Scientifica e Tecnologica, the Brazilian Conselho Nacional de
Desenvolvimento Cient\'{\i}fico e Tecnol\'ogico, CONACyT-M\'exico, the Korean
Ministry of Education, and the Korean Science and Engineering Foundation.

\end{document}